# Magnetic reversal under external field and current-driven domain wall motion in (Ga,Mn)As: influence of extrinsic pinning

K Y Wang[1], A C Irvine[2], J Wunderlich[1], K W Edmonds[3], A W Rushforth[3], R P Campion[3], C T Foxon[3], D A Williams[1] and B L Gallagher[3]

[1] Hitachi Cambridge Laboratory, Cambridge CB3 0HE, United Kingdom
Email: kw301@cam.ac.uk
[2] Microelectronics Research Centre, Cavendish Laboratory, University of Cambridge, CB3 0HE, United Kingdom
[3] School of Physics & Astronomy, University of Nottingham, NG7 2RD, United Kingdom

**Abstract.** We investigate the anisotropy of magnetic reversal and current-driven domain wall motion in annealed $Ga_{0.95}Mn_{0.05}As$ thin films and Hall bar devices with perpendicular magnetic anisotropy. Hall bars with current direction along the [110] and [1$\bar{1}$0] crystallographic axes are studied. The [110] device shows larger coercive field than the [1$\bar{1}$0] device. Strong anisotropy is observed during magnetic reversal between [110] and [1$\bar{1}$0] directions. A power law dependence is found for both devices between the critical current ($J_C$) and the magnetization (M), with $J_C \propto M^{2.6\pm0.3}$. The domain wall motion is strongly influenced by the presence of local pinning centres.

## 1. Introduction

(Ga,Mn)As, a model ferromagnetic semiconductor [1], has attracted much attention for fundamental physics and for its potential applications in spintronics [2,3]. Its magnetic anisotropy is dominated by magnetocrystalline effects which are dependent on carrier density and strain, in good agreement with theory [2]. (Ga,Mn)As epilayers grown on a relaxed (001) (In,Ga)As buffer layer experience a tensile strain due to the difference in lattice constant in each layer. Under these conditions the magnetic easy axis is perpendicular to the plane [4]. Stripe domain patterns in (Ga,Mn)As with perpendicular magnetic anisotropy have been observed previously using scanning Hall probe microscopy [5] and polar magneto-optical Kerr effect microscopy (PMOKM) [6,7,8,9]. The stripe domains are formed with a typical width of a few microns at low temperatures [5], and may be influenced by low temperature annealing [6,7,9].

It has become clear that implementation of spintronics for memory applications requires the ability to manipulate the magnetic state of a material through the application of electric fields. Manipulation of magnetic domain walls using a spin-polarized current offers a key route to this. Current-driven domain wall motion in both ferromagnetic metals and semiconductors has been demonstrated, but the mechanism is still under debate [10,11,12]. The critical current density for domain wall motion is predicted to be proportional to the saturation magnetization [13,14], which is typically two orders of magnitude smaller in (Ga,Mn)As than that in transition metal ferromagnets. The heating effect and the Oersted field produced by the critical current is correspondingly lower, so that (Ga,Mn)As is one of the best candidates for understanding current-driven domain wall motion.

Previous studies of current-driven domain wall motion in (Ga,Mn)As obtained critical current densities of around $10^5$ A/cm$^2$, which is much smaller than typically reported values for metal films. These studies were performed on films of thickness around 25 nm, with either in-plane [15] or perpendicular-to-plane easy magnetic axes [16]. In contrast, for studies of thicker (150 nm) films with in-plane magnetic easy axes, no evidence of current-driven domain wall motion was observed [17].

In the present work, we concentrate on the domain images in magnetic reversal and current-driven domain wall motion in (Ga,Mn)As with perpendicular magnetic anisotropy. In section 2, we study the magnetic domain patterns and domain wall motion under external magnetic field in the (Ga,Mn)As thin film and in patterned Hall bar devices with current oriented along [110] and [1$\bar{1}$0] crystalline axes, using PMOKM. In section 3, we combine PMOKM and magnetotransport measurements to study the current-driven domain wall motion in the (Ga,Mn)As Hall bar devices. Section 4 gives a summary of the key results of the work.

## 2. Magnetic images during magnetization reversal

The 25 nm thick $(Ga_{0.95}Mn_{0.05})$As thin film was grown on a semi-insulating GaAs (001) substrate by Molecular Beam Epitaxy using a modified Varian GEN-II system [18]. A 100 nm thick GaAs buffer layer at 580 °C, followed by a 580 nm $In_{0.15}Ga_{0.85}$As layer at 500 °C, were deposited prior to the growth of the (Ga,Mn)As layer at 255 °C. Post-growth annealing was performed in air at 190 °C for 120 hours, which is an established procedure for increasing the Curie temperature ($T_C$) of (Ga,Mn)As thin films [19]. The resulting film has Curie temperature 137±2 K, and shows very square magnetic hysteresis loops for the whole temperature range, demonstrating that the magnetic easy axis is perpendicular to the plane up to $T_C$. The PMOKM images were obtained using a commercial system with a high pressure Hg lamp and a high resolution CCD camera with time resolution up to 30 ms, giving a spatial resolution of 1 μm with image as large as 150 μm × 150 μm.

### 2.1. Magnetization reversal in thin film

The Kerr rotation angle, averaged over the image area, is proportional to the component of the magnetization pointing perpendicular to the plane of the film. Figure 1a-e shows successive PMOKM snapshots of magnetic reversal under external magnetic field at T = 90 K in the annealed (Ga,Mn)As thin film. Initially, the film is saturated with a negative magnetic field of -300 Oe, which is much larger than the coercive field. The field is then swept to +27.5 Oe, just less than the coercive field. PMOKM images are then captured at a rate of 15 frames/s. The time-resolved domain images captured during the nucleation and propagation of domain walls are shown in figures 1a to 1e. These images reveal that the domain walls are nucleated at the side of the films and align along the [1$\bar{1}$0] direction, propagating rapidly along the [110] axis between pinning sites, until the magnetization is almost fully reversed with only a few unreversed stripe domains remaining. Similar images of magnetization reversal are observed over the whole temperature range up to $T_C$. The width of the unreversed stripe domains is a few microns, which is much wider than the typical domain wall width (~15 nm) in (Ga,Mn)As [20]. The anisotropic magnetization reversal process may be attributed to an anisotropy of residual pinning sites after annealing [21], or to uniaxial in-plane magnetocrystalline anisotropy [22].

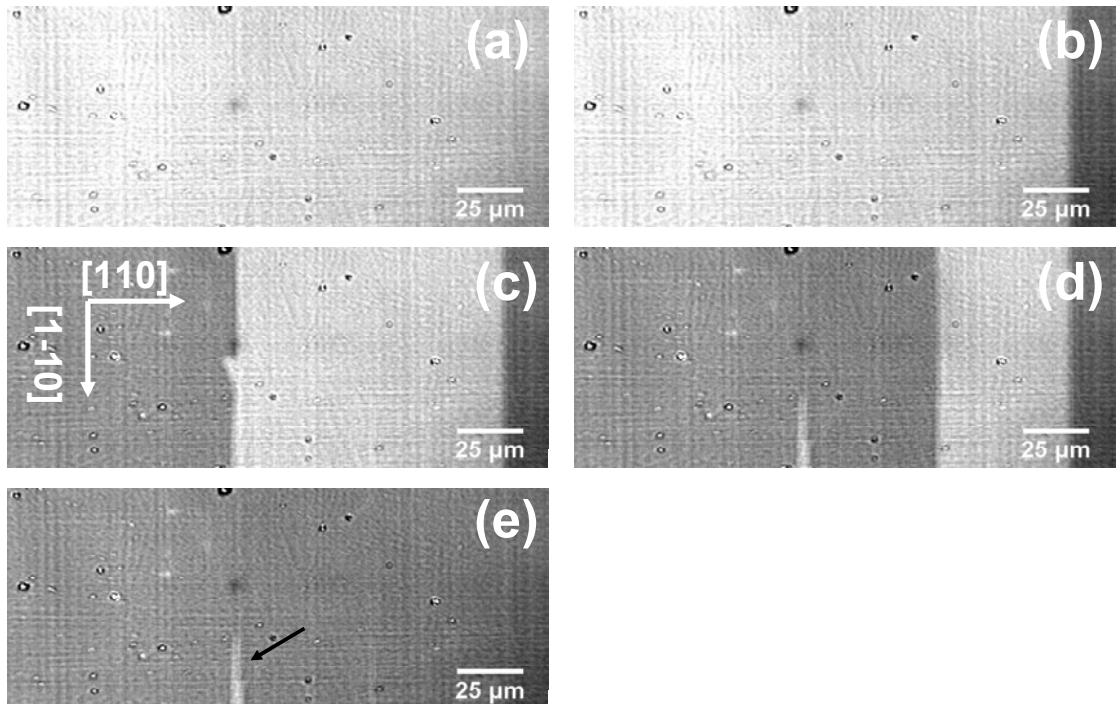

**Figure 1.** Successive PMOKM snapshots of the magnetic domain pattern during the magnetization reversal at 90K for the annealed (Ga,Mn)As thin film at H = 23.7 Oe, after (a) 5 s; (b) 5.6 s; (c) 7.3 s; (d) 9.23 s; and (e) 9.37s, respectively. The arrow in (e) indicates a persistent residual stripe domain.

### 2.2. Magnetic reversal in fabricated Hall bars

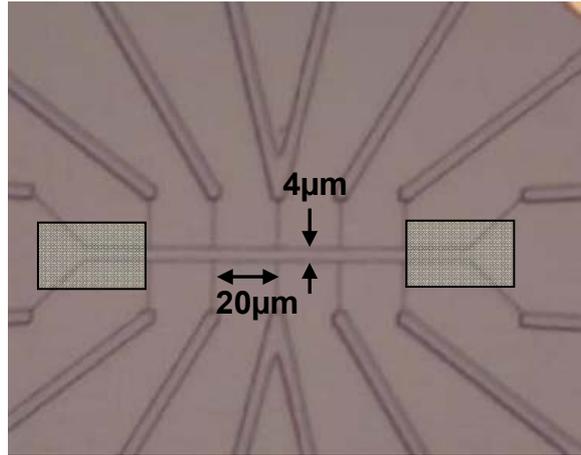

**Figure 2.** Optical image of a fabricated (Ga,Mn)As Hall bar device. The top ~10nm of the two marked rectangles have been etched away.

4μm wide modified Hall bar devices as shown in figure 2, with the current channel along either the [110] or [1$\bar{1}$0] crystalline axis, were fabricated by electron beam lithography. The length between neighbouring arms along the bar is 20 μm and the total length of the bar is 120 μm. A 10-15 nm surface layer has been etched away at both ends, as marked in figure 2. The devices were annealed in air at 190 °C for 24 hours. The Curie temperature for the [110] and [1$\bar{1}$0] devices are determined by using PMOKM to be 120±2 K and 122±2 K, respectively. The small difference in $T_C$ is due to small differences in growth parameters across the wafer. The Curie temperature of the devices is lower than that of the annealed film studied in section 2.1 due to the shorter anneal time.

The magnetic hysteresis loops are measured using PMOKM for both devices at different temperatures. During the magnetization reversal, clear domain wall propagation along the bar is observed in the [110]-oriented device. The device is first saturated in positive field, and then the field is swept to negative values at a rate of 1 Oe/s. The magnetic configuration during the domain wall propagation is shown in figure 3a-d, which corresponds to points A-D marked in the hysteresis loop in figure 3e. The domain walls can propagate from both ends of the bar, which is consistent with the magnetic reversal results for the thin film. States with pinned magnetic domain walls are observed in figure 3b and figure 3c.

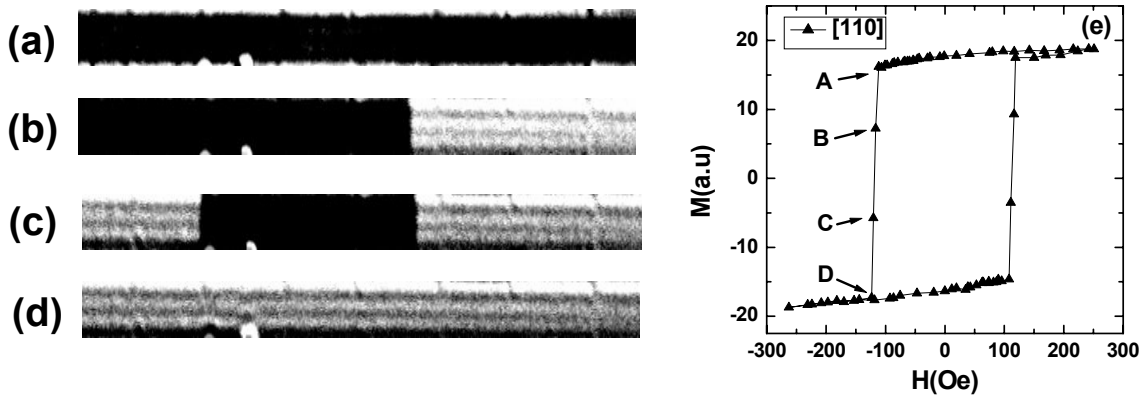

**Figure 3.** (a-d) PMOKM domain images during magnetization reversal of the [110]-oriented device at 102 K; (e) magnetic hysteresis loop for the device, with the points marked A-D corresponding to the domain configurations a-d, respectively.

Similar experiment is performed for the [1$\bar{1}$0]-oriented device. The magnetic configuration during the magnetic reversal is shown in figure 4a-c. No domain wall propagation is observed in this case, even with time resolution down to 30 ms. The magnetic images are quite homogeneous. For figure 4a and figure 4c, the magnetization is fully perpendicular to the plane (up and down, respectively), while the magnetization in figure 4b is close to zero. The latter may be because the magnetization reversal occurs on a faster timescale than the image integration time of the PMOKM measurement. This indicates that the magnetic field driven domain wall propagation across the Hall bar is much faster than our time resolution.

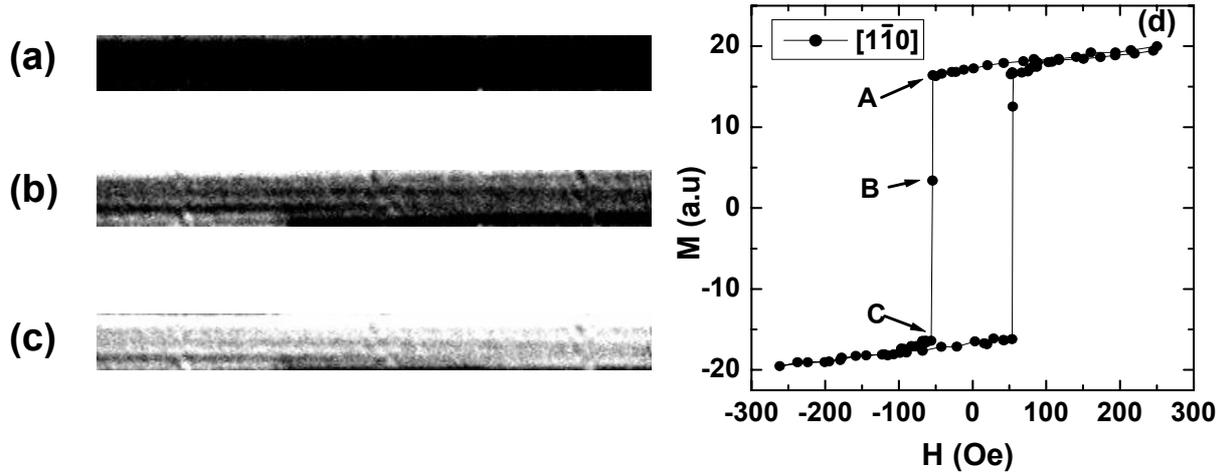

**Figure 4.** (a-c) PMOKM magnetic images of the magnetization reversal of the $[1\bar{1}0]$-oriented device at 102 K; (d) magnetic hysteresis loop for the device, with the points marked A-C corresponding to the domain configurations a-c, respectively.

The temperature dependence of the coercive field ($H_C$) obtained by PMOKM for both devices is shown in figure 5. Below 117K, $H_C$ for the $[1\bar{1}0]$-oriented device is much smaller than for the [110]-oriented device. The results are consistent with the magnetic reversal images in the unpatterned film (figure 1) which show that the domain wall propagate along the [110] direction. The coercive field for the $[1\bar{1}0]$-oriented device linearly decreases with increasing temperature, while the decrease is nonlinear for the [110]-oriented device.

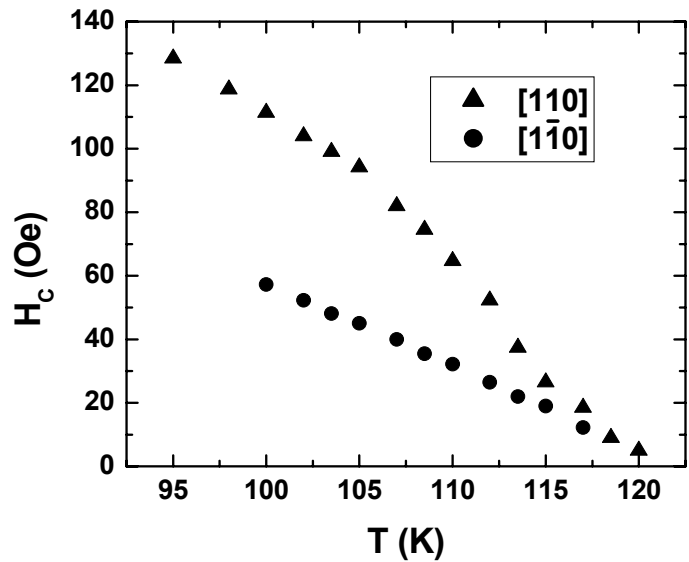

**Figure 5.** Temperature-dependence of the coercive field for the [110] and $[1\bar{1}0]$-oriented devices.

## 3. Current driven domain wall motion

In this section we discuss current-driven domain wall motion in the [110] and [1$\bar{1}$0]-oriented devices. Due to the different coercive field for the etched and non-etched parts of the devices, we can initialize a magnetic configuration with domain walls formed at both interfaces between etched and non-etched regions, using an external magnetic field. We then determine the critical current $J_C$ required to move the domain wall from the interface. Irrespective of the relationship between the applied current direction and the domain wall direction, we expect to observe domain wall motion within the PMOKM image window.

*3.1. DC current-driven domain wall motion*

For the current-driven domain wall studies, we initially form domain walls at both interfaces using the external magnetic field, and then reduce the field to zero and also screen the light in order to prevent the complication of photoexcited effects. Increasing the dc current from zero, we monitor the Hall resistance for both the A and B Hall crosses simultaneously (see figure 6). Local reversal of the magnetization is detected as a large change in the Hall signal, due to the anomalous Hall Effect. When any abrupt change of the Hall resistance from either pair of contacts is observed, the dc current is switched to zero, and the magnetic configuration is imaged using PMOKM.

In order to distinguish spin-transfer induced domain wall motion from Joule heating or Oersted field induced effects, we investigate four different configurations of magnetization and applied current direction, illustrated in figures 6 and 7:

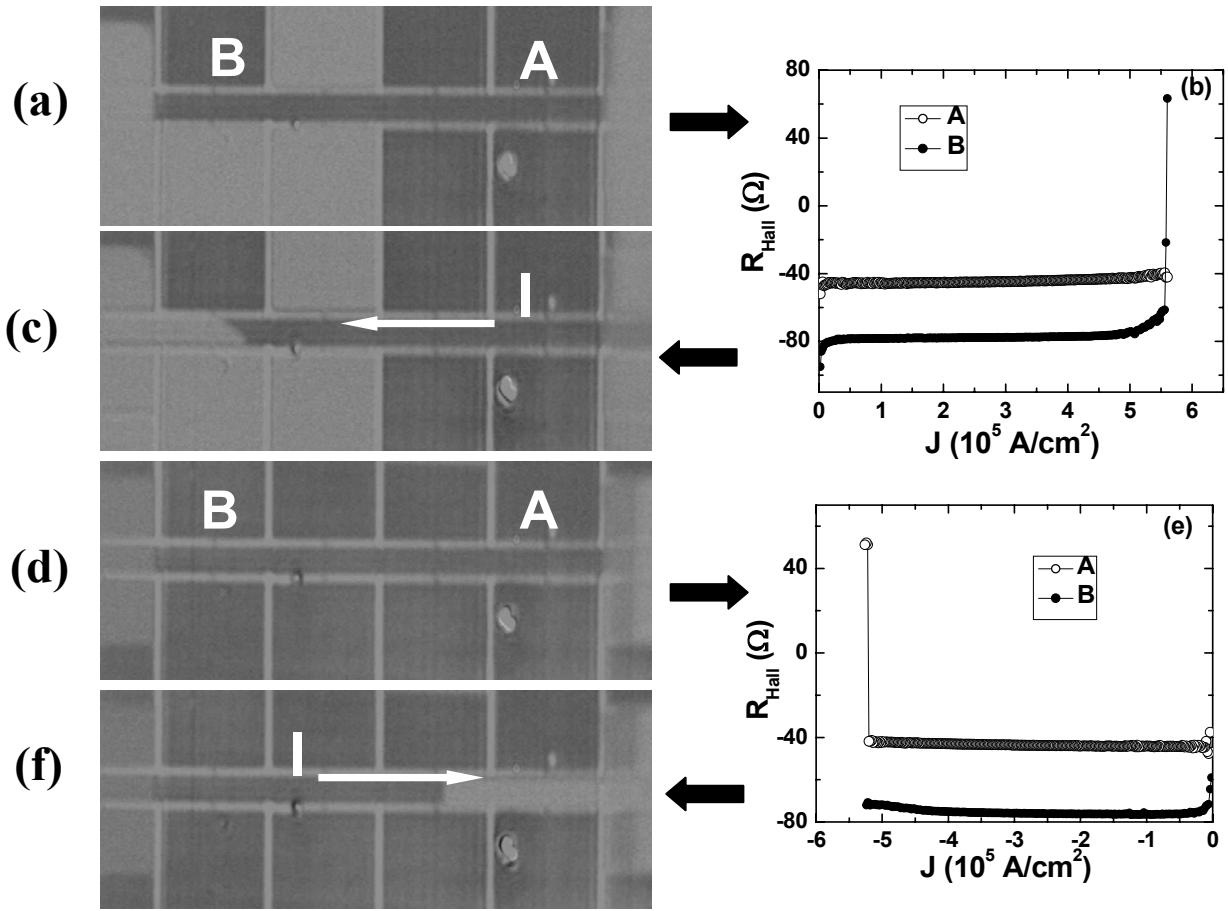

**Figure 6.** (a-c) dc current driven domain wall motion with positive electric current: (a) the initial magnetic configuration with domain walls at each interface; (b) the *in-situ* monitored Hall resistance at crosses A and B during application of positive dc current; (c) magnetic configuration after applying the positive current shown in (b). (d-f) dc current driven domain wall motion with negative electric current: (d) initial magnetic configuration; (e) *in-situ* monitored Hall resistance at crosses A and B during application of negative dc current; (f) magnetic configuration after applying the negative current shown in (e). The arrows in (a), (c), (d), (f) indicate the current direction.

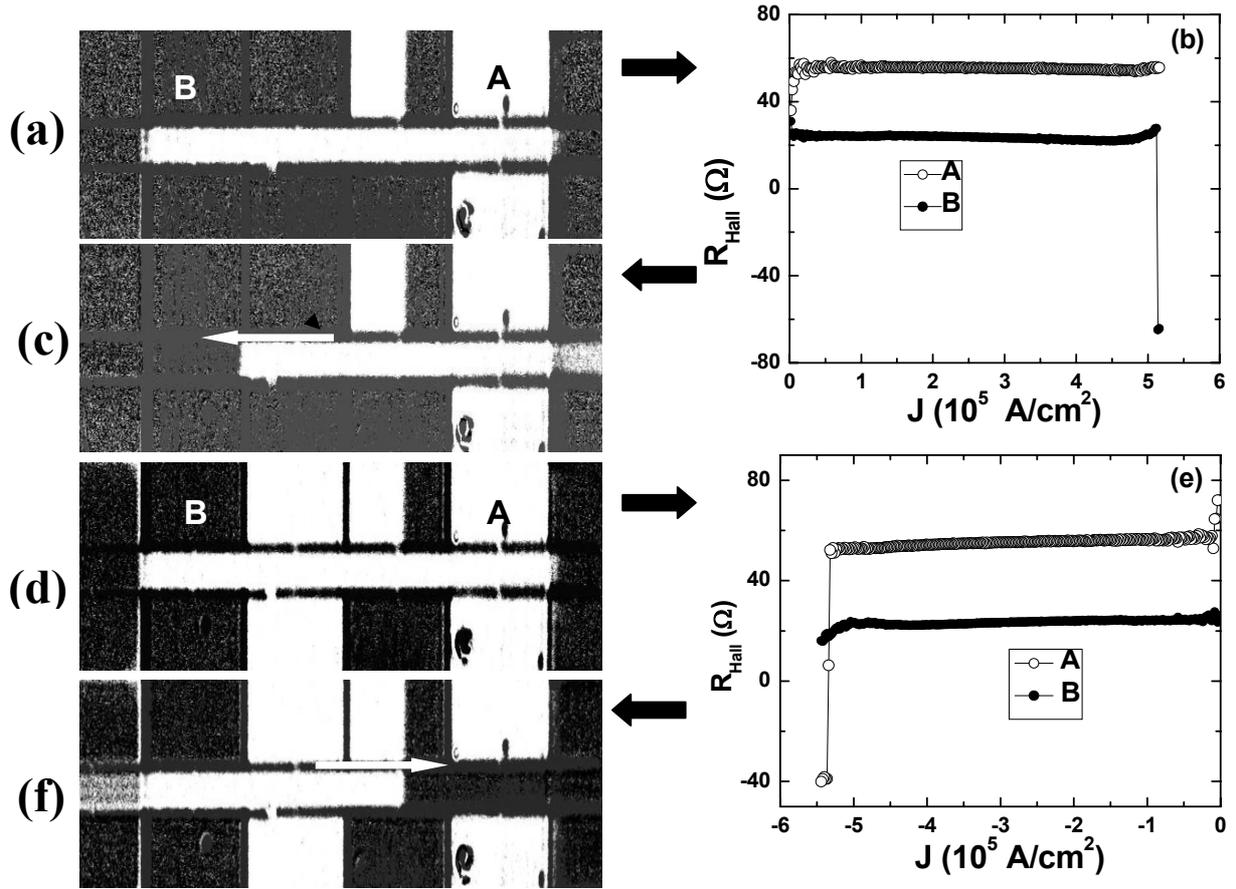

**Figure 7.** dc current driven domain wall motion as in figure 6, but with opposite initial magnetic configuration: (a) the initial magnetic configuration with domain walls at each interface; (b) the *in-situ* monitored Hall resistance at crosses A and B during application of positive dc current; (c) magnetic configuration after applying the positive current shown in (b); (d) initial magnetic configuration; (e) *in-situ* monitored Hall resistance at crosses A and B during application of negative dc current; (f) magnetic configuration after applying the negative current shown in (e).

i) Initially the magnetization is saturated with a positive magnetic field of 1000 Oe, which is much higher than the coercive field. The field is then swept to -105 Oe, which switches only the etched regions of the device, with domain walls formed at each interface. The magnetic field is then reduced to zero. The image of this initial magnetic configuration for the [110]-oriented device at T=102 K is shown in figure 6a. A positive electric current is then applied from zero at a rate of $2 \times 10^3$ A/cm$^2$s. The monitored Hall resistance at crosses A and B versus increasing current is shown in figure 6b. A sharp change of the Hall resistance at cross B is observed when the current reaches $\sim 5.6 \times 10^5$ A/cm$^2$. We define this onset point of rapidly changing anomalous Hall resistance as the critical current density $J_C$ for dc current-driven domain wall motion. The magnetic configuration image after applying the positive dc current, shown in figure 6c, indicates that the left side domain wall has propagated around 15 μm along the channel towards the right side. Therefore, the direction of motion of the domain wall is opposite to the electrical current direction.

ii) The magnetization configuration is initialized as before (figure 6d), but this time a negative dc current is applied from zero at a rate of $-2 \times 10^3$ A/cm$^2$s. The resulting Hall resistance versus increasing negative current for crosses A and B is shown in figure 6e. A sharp change of the Hall resistance of cross A is observed. The magnetic configuration image after applying the negative dc current (figure 6f) shows that the right domain wall has propagated around 30μm along the stripe towards the left side. Therefore, the domain wall moves in the opposite direction to the current as before, with similar $J_C$.

iii) and iv) The magnetization is saturated with a negative magnetic field of -1000 Oe, before the field is swept to +105 Oe and then zero. This yields the opposite magnetic configuration to figures 6a and 6d, as shown in figures 7a and 7d. The Hall resistance is shown against positive and negative dc currents in figure 7b and 7e, respectively, with the current swept from zero at a rate of $\pm 2\times 10^3$ A/cm$^2$s as before. The resulting magnetic configuration images after the critical current is reached are shown in figure 7c and 7f. In both cases, the domain wall motion is again opposite to the current direction.

The above observations rule out the Oersted field or Joule heating as the origin of the current-driven domain wall motion. Similar current-induced domain wall motion is observed for both the [110]- and [1$\bar{1}$0]-oriented devices. Our results demonstrate that the spin-transfer torque is indeed the origin of the current-driven domain wall motion in the present devices. The sign and magnitude of the critical current for domain wall motion is in agreement with previous studies of (Ga,Mn)As devices with perpendicular magnetic anisotropy [12,16].

*3.2. Single current pulse driven domain wall motion*
In order to minimize the heating effect associated with the dc current, we also investigate domain wall motion induced by single current pulses of width 1 ms. We keep the same initial magnetic configuration as in the dc current measurements and image the initial state. The light is screened while the current pulse is applied. Then the final magnetic configuration is imaged after applying a current pulse of varying current density. The critical current density is then defined as the value at which domain motion occurs, as identified by a difference between initial and final magnetic configurations.

*3.3. Critical current for domain wall motion*
Before correcting the device temperature for current-induced heating effects, the critical current density obtained from dc current measurements is lower than that of pulsed current measurements. The difference of $J_C$ obtained by these two methods becomes larger as the device temperature is reduced. This is due to the larger heating effect in the dc measurements. In order to take into account the heating effect of the electric current, we use the longitudinal resistance of the device to calibrate the device temperature. The longitudinal resistance during the current-driven domain wall motion is complicated by the presence of anisotropic magnetoresistance and anomalous Hall resistance. Therefore, we cannot directly use the longitudinal resistance obtained from dc measurements and single current pulse measurements to calibrate the device temperature. Instead, the calibration was performed under an external magnetic field of 300 Oe. This ensures that the magnetic configuration is unchanged even with large dc current applied. We compare the longitudinal resistance for the device at low current density $J=8\times 10^3$ A/cm$^2$ with the longitudinal resistance obtained for dc currents applied from zero at a rate of $\pm 2\times 10^3$ A/cm$^2$ and for single current pulses. The heating effect from the dc current (at critical current) and pulsed current is calibrated separately for both devices as a function of temperature.

The corrected temperature-dependence of the critical current density for current-driven domain wall motion for the [110]- and [1$\bar{1}$0]-oriented devices is shown in figure 8a. After calibration of the device temperature, the critical currents obtained from dc current and pulsed current measurements are consistent with one another. The critical current decreases with increasing temperature for both devices, due to the weakening of the magnetization and magnetic anisotropy as the Curie temperature is approached. Between 105 K and 116 K, significant differences are observed between the critical current obtained for the different oriented devices. We attribute this to the small difference in $T_C$ and rapid change of the magnetization close to $T_C$.

The temperature-dependence of the saturation magnetization obtained from PMOKM for both devices is shown in figure 8b. This is in good agreement with the Brillouin function for a ferromagnet with *S*=5/2, with $T_C$=120 K and 122 K respectively. Using the Brillouin function to obtain the normalized magnetization, the critical current density $J_C$ versus magnetization for both devices is shown in a log-log plot in figure 8c. The critical current density follows the same power law dependence on the magnetization for both devices within the experimental error, with $J_C \propto M^{2.6\pm 0.3}$. Therefore, the critical current decreases more rapidly with decreasing M than the linear relationship given by existing theories [14]. This may be due to a strong magnetization-dependence of the domain wall pinning at the etch step.

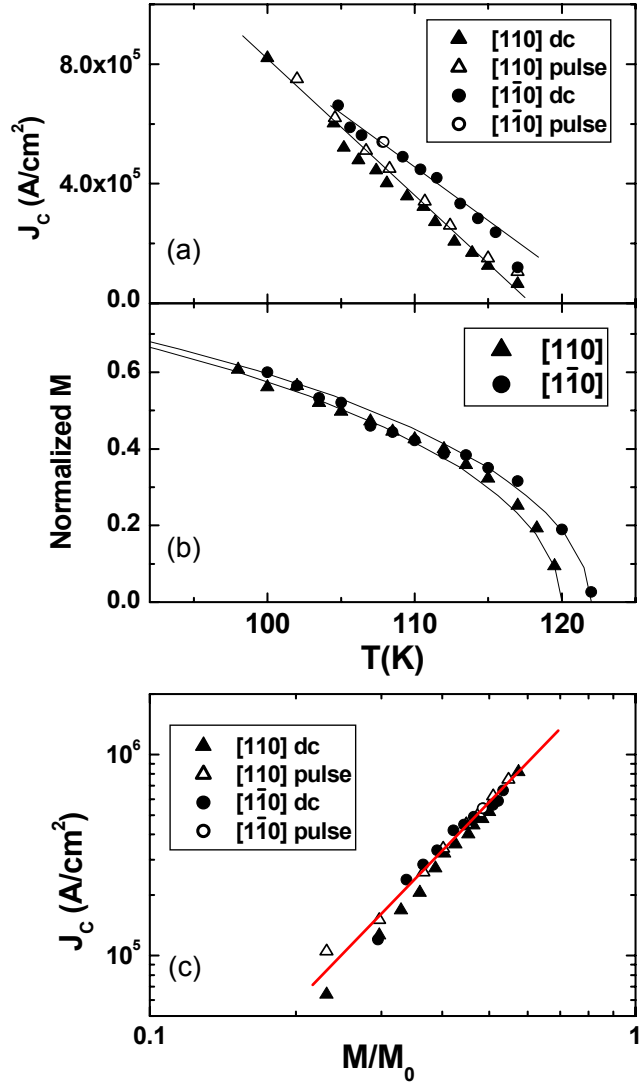

**Figure 8.** (a) Temperature dependence of the critical current density obtained from dc measurements (solid symbols) and single pulse measurements (open symbols) for the two devices, with lines to guide the eye. (b) Temperature dependence of the normalized magnetization for the two devices obtained by PMOKM, with lines showing the Brillouin functions with *S*=5/2 for Curie temperatures of 120K and 122K. (c) Magnetization dependence of the critical current obtained from dc measurements (solid symbols) and single pulse measurements (open symbols) for the two devices, with the line showing the power law fit.

*3.4. Domain wall displacement in (Ga,Mn)As*
We next investigate the domain wall displacement induced by a single current pulse of fixed width 1 ms and varying amplitude. Domain wall displacements induced by single current pulses with varying density are shown in figure 9 for the [110]-oriented device and figure 10 for the [1$\bar{1}$0]-oriented device. The device temperatures during these measurements are 102 K and 107 K, respectively. For the [110]-oriented device, the magnetic domain wall remains pinned at the interface between etched and non-etched regions for pulsed current density J=6.0x10$^5$ A/cm$^2$, as shown in figure 9a. At J=6.2x10$^5$ A/cm$^2$, domain wall motion occurs, as the domain wall has moved away from the interface in the opposite direction to the current pulse (figure 9b). The domain wall now has a wedge-like profile, which may be attributable to the non-uniformity of the current distribution around the wall [16,23], and the Oersted field produced by the current. For J=6.75x10$^5$ A/cm$^2$, the domain wall moves further from the interface, and the wedge-like distortion of the wall increases (figure 9c). However, for J=8x10$^5$ A/cm$^2$, we observe a sharp domain wall aligned perpendicular to the current direction (figure 9d). Since neither the Oersted field nor the non-uniformity of the current distribution should decrease with increasing current, we attribute this to a strong pinning line along the $[1\bar{1}0]$ axis, where a much larger critical current is required to depin the domain wall. In contrast, for the $[1\bar{1}0]$-oriented device, with increasing current density the domain wall becomes monotonically more distorted as it progresses along the bar (figure 10a-e), in agreement with previous findings [16,23].

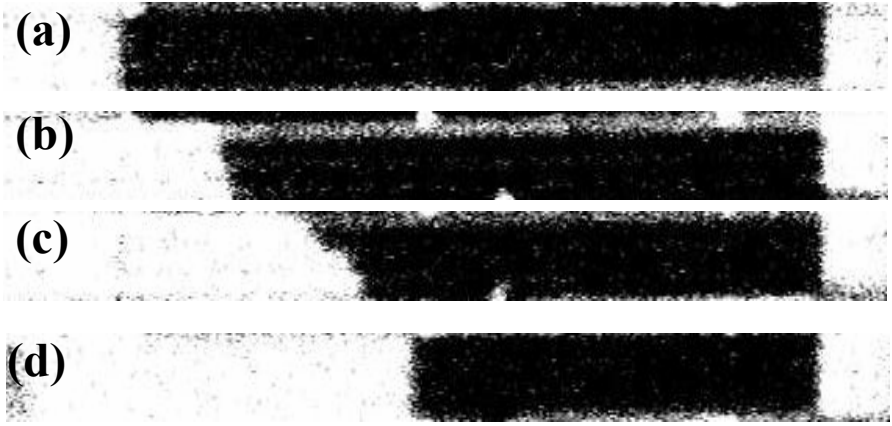

**Figure 9.** PMOKM images for the [110]-oriented device at 102 K, showing the domain wall position and profile after applying single current pulses of current density (a) $6.0\times10^5$ A/cm$^2$, (b) $6.2\times10^5$ A/cm$^2$, (c) $6.75\times10^5$ A/cm$^2$, (d) $8.0\times10^5$ A/cm$^2$.

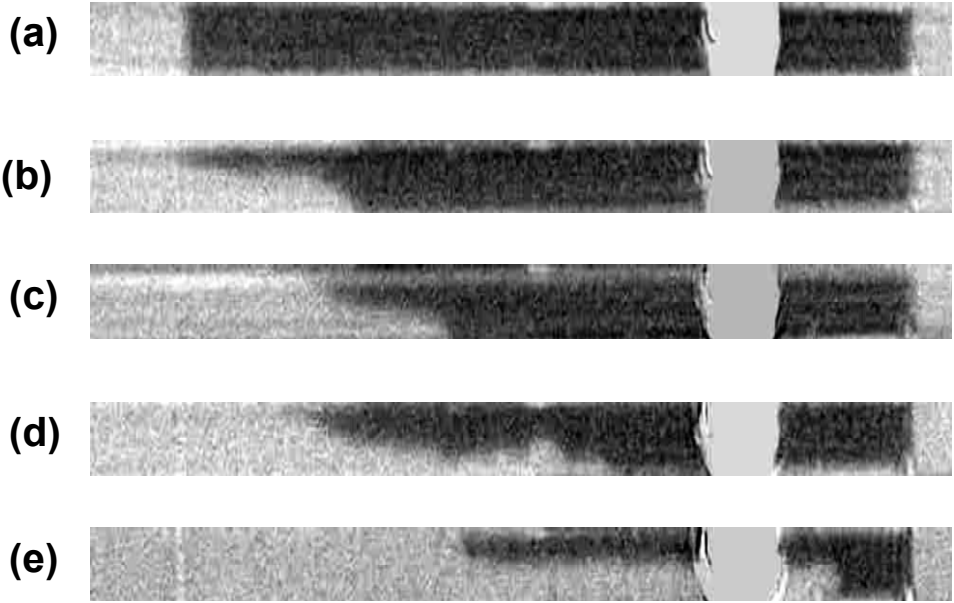

**Figure 10.** PMOKM images for the $[1\bar{1}0]$-oriented device at 107 K, showing the domain wall position and profile after applying single current pulses of current density (a) $5.2\times10^5$ A/cm$^2$, (b) $5.4\times10^5$ A/cm$^2$, (c) $5.6\times10^5$ A/cm$^2$, (d) $5.8\times10^5$ A/cm$^2$, (e) $6\times10^5$ A/cm$^2$.

The presence of strong pinning lines along the $[1\bar{1}0]$ axis is further evidenced by figure 11 and the accompanying movie, which shows PMOKM images taken at 100 K for the [110]-oriented device, during ramping the dc current from $4.5\times10^5$ A/cm$^2$ to $7.0\times10^5$ A/cm$^2$ at a rate of $\sim2\times10^3$ A/cm$^2$. Initially, the domain wall is formed at the etch step using the external magnetic field, which is then reduced to zero as the current is applied. In figure 11b and figure 11c, the wedge domain wall profile is observed. In figure 11d, the domain wall is pinned along the $[1\bar{1}0]$ axis. The domain wall remains at this pinning site as the current density is further increased, until J reaches $6.58 \times 10^5$ A/cm$^2$, where the domain wall moves rapidly along the bar until it reaches another pinning site (figure 11e). A large critical current is required to release the domain wall from the pinning sites shown in figure 11c and figure 11e.

The dependence of the domain wall position on the pulsed current density, for the two devices at different temperatures, is shown in figure 12. For the [110]-oriented device, the domain wall shows a step-like displacement as the current density is increased, as it moves rapidly between a series broad plateaus representing the strong pinning lines aligned along the $[1\bar{1}0]$ axis. The slope is much steeper for the $[1\bar{1}0]$-oriented device and the plateau regions are much narrower, indicating that the pinning sites preventing domain wall motion along the $[1\bar{1}0]$ axis are relatively weaker. The strong domain wall pinning in our devices prevents determination of the domain wall velocity in response to a current pulse, as the domain wall does not move smoothly during the pulse duration, and the domain wall displacement is mainly determined by the current density rather than the pulse width.

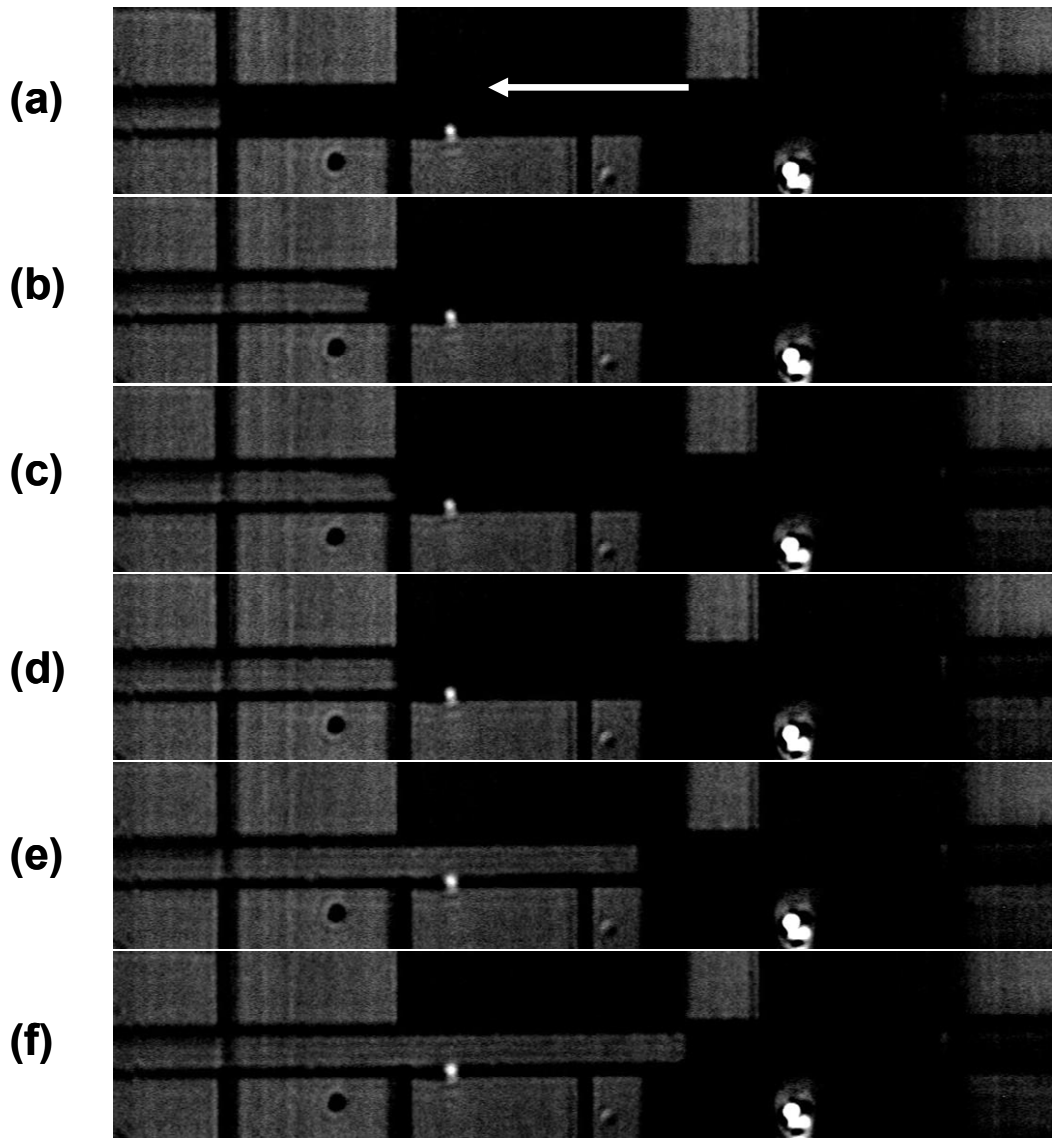

**Figure 11.** Snapshots of PMOKM movie of the [110]-oriented device obtained while increasing the dc current density at 100 K from $4.5 \times 10^5$ A/cm$^2$ at a rate of $2 \times 10^3$ A/cm$^2$s, to (a) $5.02 \times 10^5$ A/cm$^2$, (b) $5.03 \times 10^5$ A/cm$^2$, (c) $5.05 \times 10^5$ A/cm$^2$, (d) $6.57 \times 10^5$ A/cm$^2$, (e) $6.58 \times 10^5$ A/cm$^2$, (f) $7.00 \times 10^5$ A/cm$^2$. See accompanying multimedia file (the current density is shown at left-up with unit $10^5$ A/cm$^2$) (file size 1MB, .avi format).

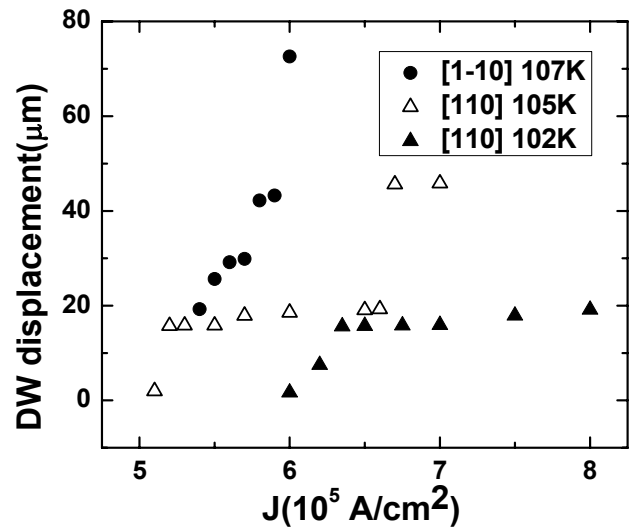

**Figure 12.** Domain wall displacement after application of a single current of varying density, for the [110]- and [1$\bar{1}$0]-oriented devices at different temperatures.

## 4. Conclusions

Using PMOKM and electrical transport measurements we have investigated domain wall propagation in Hall bars and thin films of tensile strained (Ga,Mn)As with perpendicular magnetic anisotropy, in response to applied magnetic fields and spin polarized electric currents. The anisotropy of domain wall motion and domain wall pinning sites observed for (Ga,Mn)As films grown on strain-relaxed (In,Ga)As [7,9] leads to a number of important differences in the behaviour of Hall bars with current channel along the [110] and [1$\bar{1}$0] in-plane crystalline axes. A much smaller coercive field is observed during magnetization reversal for the [110]-oriented device compared to the [1$\bar{1}$0]-oriented device. In addition, strong pinning lines are observed in the [110]-oriented device, resulting in domain walls oriented perpendicular to the current direction, with large associated critical currents. Therefore, the domain wall displacement induced by a pulsed current in these devices is mainly determined by the current density, rather than the width of the pulse. However, for a domain wall trapped at an etch step, similar critical currents are obtained for the [110]- and [1$\bar{1}$0]-oriented device, and a power law dependence is observed between the critical current and the temperature-dependent magnetization, given by $J_C \propto M^{2.6\pm0.3}$.

## 5. Acknowledgement
The authors thank G. Tatara for valuable discussions. This project was supported by EC sixth framework grant FP6-IST-015728.

**References**
[1] Jungwirth T, Sinova J, Masek J, Kucera J and MacDonald A H 2006 *Rev. Mod. Phys.* **78**, 809.
[2] Dietl T, Ohno H and Matsukura F 2001 *Phys. Rev. B* **63**, 195205.
[3] Wolf S A, Awschalom D D, Buhrman R A, Daughton J M, Von Molnar S, Roukes M L, Chtchlkanova A Y and Treger D M 2001 *Science* **294**, 148.
[4] Shen A, Ohno H, Matsukura F, Sugawara Y, Akiba N, Kuroiwa T, Oiwa A, Endo A, Katsumoto S and Iye Y 1997 *J. Crystal Growth* **175/176**, 1069.
[5] Shono T, Hasegawa T, Fukumura T, Matsukura F and Ohno H 2000 *Appl. Phys. Lett.* **77**, 1363.
[6] Thevenard L, Largeau L, Mauguin O, Patriarche G, Lemaitre A, Vernier N and Ferre J 2006 *Phys. Rev. B* **73**, 195331.
[7] Dourlat A, Jeudy V, Testelin C, Bernardot F, Khazen K, Gourdon C, Thevenard L, Largeau L, Mauguin O and Lemaitre A 2007 *J. Appl. Phys.* **101**, 106101.
[8] Chiba D, Matsukura F and Ohno H 2006 *Appl. Phys. Lett.* **89**, 162505.
[9] Wang K Y, Rushforth A W, Grant V A, Campion R P, Edmonds K W, Foxon C T, Gallagher B L, Wunderlich J and Williams D A 2007 *J. Appl. Phys.* **101**, 106101.

[10] Fert A, Cros V, George J-M, Grollier J, Jaffrès H, Hamzic A, Vaurès A, Faini G, Youssef J B, and LeGall H 2004 *J. Magn. Magn. Mater.* **272**, 1706.
[11] Hayashi M, Thomas L, Rettner C, Moriya R and Parkin S S P 2007 *Nature Phys.* **3**, 21.
[12] Yamanouchi M, Chiba D, Matsukura F and Ohno H 2004 *Nature (London)* **428**, 539.
[13] Zhang S and Li Z 2004 *Phys. Rev. Lett.* **93**, 127204.
[14] Tatara G and Kohno H 2004 *Phys. Rev. Lett.* **92**, 086601.
[15] Wunderlich J *et al.* 2007 *Phys. Rev. B* **76**, 054425.
[16] Yamanouchi M, Chiba D, Matsukura F, Dietl T and Ohno H 2006 *Phys. Rev. Lett.* **96**, 096601.
[17] Tang H X, Kawakami R K, Awschalom D D and Roukes M L, 2006 *Phys. Rev. B* **74**, 041210(R).
[18] Campion R P, Edmonds K W, Zhao L X, Wang K Y, Foxon C T, Gallagher B L and Staddon C R 2003 *J. Crystal Growth* **247**, 42.
[19] Edmonds K W, Wang K Y, Campion R P, Neumann A C, Farley N R S, Gallagher B L and Foxon C T 2002 *Appl. Phys. Lett.* **81**, 4991.
[20] Dietl T, Konig J and MacDonald A H 2001 *Phys. Rev. B* **64**, 241201(R).
[21] Maksimov O, Sheu B L, Xiang G, Keim N, Schiffer P and Samarth N 2004 *J. Crystal Growth* **269**, 298.
[22] Liu X, Sasaki Y and Furdyna J K 2003 *Phys. Rev. B* **67**, 205204.
[23] Yamanouchi M, Ieda J, Matsukura F, Barnes S E, Maekawa S and Ohno H 2007 *Science* **317**, 1726.